\documentclass{jfm}
\usepackage{graphicx}
\usepackage{epstopdf, epsfig}
\usepackage{xcolor}
\usepackage{siunitx}

\usepackage{amsmath}
\usepackage{mathtools} 
\usepackage{bm}
\usepackage{subfigure}
\usepackage{cancel}
\usepackage{comment}

\title{Axisymmetric predictions for mitigated and vertically unstable disruptions in ITER with runaway electrons}
\author{V. Bandaru\aff{1,2}, M. Hoelzl\aff{2}, F.J. Artola\aff{3}, M. Lehnen\aff{3}, and JOREK team}
\affiliation{ \aff{1}Indian Institute of Technology Guwahati, Assam, India
\aff{2}Max Planck Institute for Plasma Physics, Garching, Germany
\aff{3} ITER Organization, Route de Vinon sur Verdon, 13067 St Paul Lez Durance Cedex, France}

\begin{document}

\maketitle

\begin{abstract}
We present two-dimensional global simulations of mitigated and vertically unstable disruptions in ITER in the presence of runaway electrons. An elongated plasma in free-boundary equilibrium is subjected to an artificial thermal quench and current-profile flattening, followed by one or more massive material injections and RE avalanche. Scenarios of major disruptions as well as upward and downward vertical displacement events are considered. Results provide important insights into the effects of runaway electron formation, post thermal quench current-profile, injection quantities and timings, and impurity flushout on the overall evolution of disruption and the plasma vertical motion thereof. Interplay between the various effects offers scope for potentially beneficial runaway electron mitigation scenarios.
\end{abstract}

\section{Introduction}

Tokamak plasmas are known to occasionally get into states, wherein  small perturbations grow exponentially leading to a disruption or termination of the discharge \cite[]{Waddell:1979}. Such a disruptive event is typically described by various phases that the plasma undergoes. Non-linear interaction of large scale magneto-fluid instabilities cause a stochastization of the magnetic field structure, effectively causing most of the plasma thermal energy to be lost on a short timescale ($\sim 1$ms in many existing devices). Such a thermal quench (TQ) phase increases the plasma electrical resistivity by orders of magnitude, and is therefore followed by a current quench (CQ) phase wherein the plasma current decays at a timescale of resistive diffusion. During the CQ phase, depending on the plasma conditions, it is possible that a seed of high energy relativistic electrons in the plasma grows exponentially so as to form a beam that carries nearly the entire plasma current by the end of CQ \cite[]{Breizman:2019}. In fusion relevant tokamak devices, the current carried by such runaway electron (RE) beams can be a significant fraction of the pre-disruption current. Moreover, the preferred choice of vertically elongated plasmas in today's devices makes the plasma conducive to a vertical instability or a vertical motion of the plasma towards the first-wall, often referred to as a vertical displacement event (VDE). A VDE can either precede and be the primary driver of a thermal quench (a hot-VDE), or can follow the thermal quench in which case it is referred to as a cold-VDE \cite[]{Hender:2007, Artola:2024}. 
While disruptions (in whichever form they occur) aren't a serious issue in small and medium size tokamak devices, their consequences can be potentially severe in fusion-grade tokamaks such as ITER. They can manifest as one or more of potentially damaging thermal loads on the divertor, enormous electromagnetic forces on the surrounding conducting structures and RE induced thermal loads that can melt the first wall and beyond. It therefore becomes imperative to try to avoid or mitigate the effects of disruptions.

The aim of disruption mitigation is to simultaneously address all the detrimental consequences of disruptions: the thermal loads, electromagnetic loads and RE induced damage \cite[]{Eidietis:2021}. Towards this end, over the years several mitigation strategies have been proposed and evaluated for ITER. Currently, the ITER disruption mitigation system (DMS) is based on the injection of massive quantities of Neon and/or Deuterium into the plasma in the form of shattered pellets, in one or more stages \cite[]{Lehnen:2023}. Depending on the specifics of any given disruption, there can be a few different schemes of executing the injections. Nevertheless, all the schemes essentially take advantage of the following general effects of Neon and Deuterium. Neon aids in effectively radiating the thermal energy, thereby reduces divertor thermal load and also avoids very long current quench times. In the context of RE beam mitigation, while impurities can cause a faster decay of the beam, they also aid in significantly increasing the avalanche growth (an unintentional effect in spite of the increase in critical electric field) via the additional bound electrons which act as additional targets for the avalanche process. On the other hand, Deuterium during first injection can aid in plasma pre-dilution and reduced RE hot-tail seed \cite[]{Nardon:2020}, and during 2\textsuperscript{nd} injection onto a fully/partially formed RE beam, it can aid in neutralizing the ions (flushout) and cause a benign loss of REs \cite[]{Reux:2021, Hollmann:2023}.  In addition to the above effects, inherent vertical instability of an elongated plasma (as is the case with ITER) leads to a vertical motion of the plasma column while it is going through massive material injections and RE formation. It is therefore evident that any reliable estimate or assessment of a disruption must self-consistently include the effects of RE formation, vertical motion and material injection at the very least. This is the main motivation of the present work.

In this work, we present free-boundary 2D/axisymmetric simulations of an enlongated ITER-like plasma that undergoes a disruption, using the fusion MHD code JOREK \cite[]{Czarny:2008, Hoelzl:2021}. The plasma is first subjected to an artificial thermal quench and current-profile flattening, followed by a first injection of a mixture of Neon and Deuterium, alongside an introduction of an RE seed. Subsequent to this, the plasma goes through current quench and RE avalanche growth, vertical motion, and Neon 2\textsuperscript{nd} injection and/or flushout. The interplay of the various effects are studied considering different disruption types (major disruption, upward and downward VDEs), different injection quantities and timings, and different types of current-profile flattening. The effects of conducting structures surrounding the plasma are accounted for via the JOREK-STARWALL coupling \cite[]{Merkel:2006, Hoelzl:2012, Artola-thesis:2018}, while the runaway electrons are treated using an RE fluid model \cite[]{Bandaru:2019, Bandaru-model:2024} that is electromagnetically coupled to the background plasma. Effects of partially-ionized impurities on the RE avalanche are taken into account through the models of \cite{Hesslow:2018-PPCF, Hesslow:2019}. 

The paper is organized as follows. In section~2 we briefly present the physical model used for the simulations. In section~3,  the details of the simulation set-up and the various physical properties used in the simulations are presented. This is followed by a detailed analysis of the results in section~4, followed by a summary of the key conclusions and outlook in section~5.

\section{Physical model}

As mentioned earlier, for the present simulations we use the fusion MHD code JOREK. In particular, a reduced MHD version of the code is used that additionally accommodates impurities as well as runaway electrons. Individual charge states of the impurities are not evolved separately. Instead, the impurity charge state distribution is obtained using the coronal equilibrium model \cite[]{Mosher:1974}, while the code in general also has more advanced impurity models with higher computational costs \cite[]{Hu:2021}. The runaway electrons are treated as an extra fluid species separate from the background plasma and impurities. The RE fluid couples electromagnetically to the background plasma, via the modification that arises in the Ohm's law due to the REs not being subjected to resistive decay. Furthermore, only the density of the RE fluid ($n_r$) is evolved, without considering the dynamics of RE pitch angle and momentum. The RE density evolution includes an RE avalanche volumetric source along with parallel transport via a large parallel diffusivity and the $\bm{E}\times \bm{B}$ drift. Parallel diffusivity is used to mimic the effect of RE advection at the speed of light in a computationally cost effective manner (see \cite{Bandaru:2019, Bandaru-PPCF:2021} for details). The RE avalanche source includes the effects of partially-ionized impurities (i.e. the enhanced critical electric field as well as the bound electron targets for avalanche) through the models developed in \cite{Hesslow:2018-PPCF, Hesslow:2019}. However, the effect of REs on the charge-state distribution of the impurities is neglected.

The electromagnetic coupling of all the passive electrically conducting structures as well as the current carrying coils are accounted for via the JOREK-STARWALL coupling. The coupling involves using non-local electromagnetic boundary conditions for the plasma using Green's function, without physically extending the domain or grid into the vacuum region. Furthermore, for the simulations presented in this paper, the diamagnetic drift as well as the field parallel velocity of the plasma fluid are not considered to reduce computational costs for the long timescales simulated in the present work. In other words, the background plasma fluid velocity is assumed to comprise only the $\bm{E}\times \bm{B}$ drift. Comprehensive details of the reduced MHD model as well as the numerical schemes involved in JOREK are described in \cite{Czarny:2008, Hoelzl:2021}, while the specifics of the JOREK-STARWALL coupling can be found in \cite{Merkel:2006, Hoelzl:2012, Artola-thesis:2018}. The RE fluid model used hereby (along with model benchmark studies) have been described in detail in \cite{Bandaru-model:2024}.

\section{Overview of the simulation setup }

\begin{figure}
\subfigure[]{
\centering
\includegraphics[trim = 0mm 0mm 0mm 0mm, clip, width=0.6\textwidth]{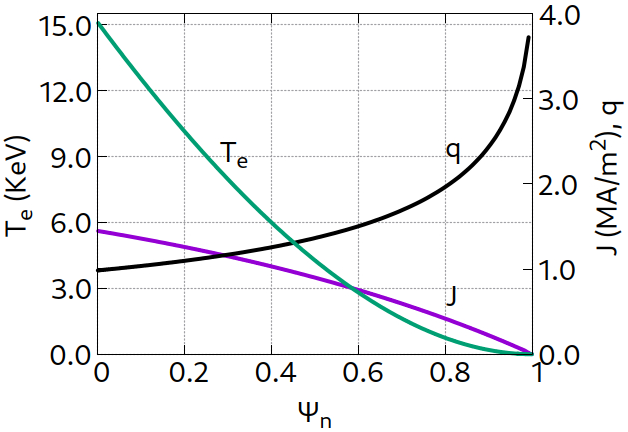}
\label{fig:equil_Te_profile}
}
\hspace{0.3cm}
\subfigure[]{
\centering
\includegraphics[trim = 195mm 30mm 170mm 50mm, clip, width=0.35\textwidth]{equil_flux1.pdf}
\label{fig:equil_flux_surfaces}
}

\hspace{1.5cm}
\subfigure[]{
\centering
\includegraphics[trim = 0mm 0mm 0mm 0mm, clip, width=0.75\textwidth]{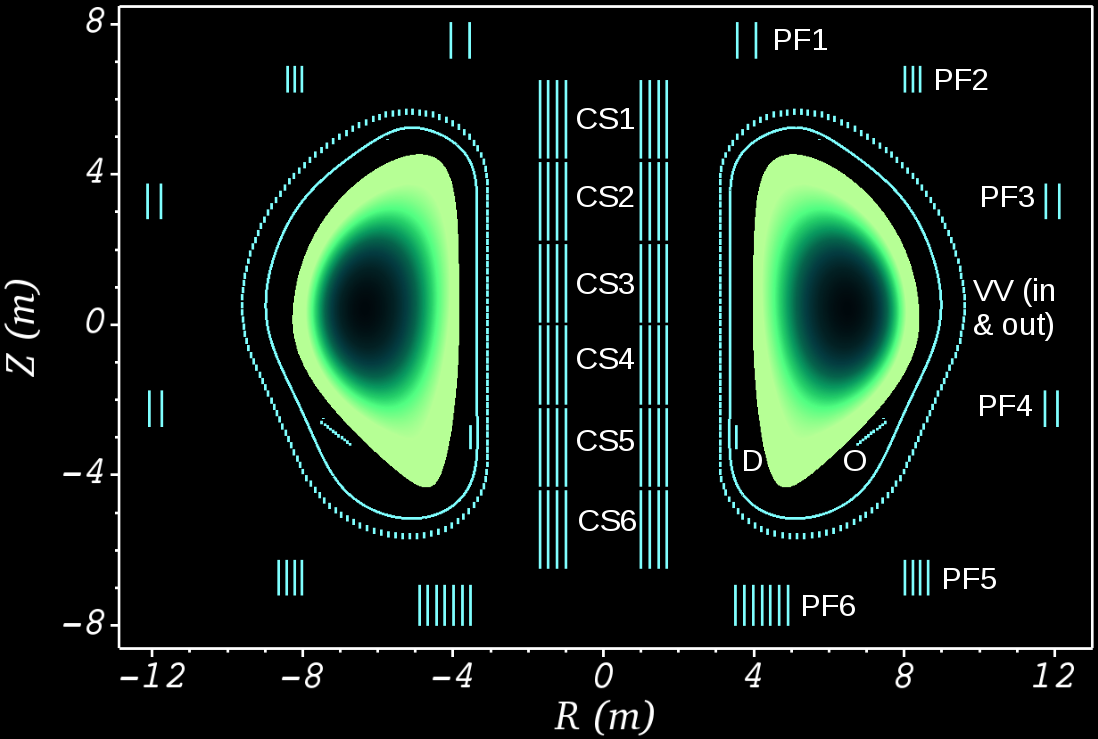}
\label{fig:cond_structures}
}
\caption{a) Initial equilibrium plasma profiles of electron temperature $T_e$ , toroidal current density ($J$), the safety factor ($q$) as a function of normalized poloidal magnetic flux $\psi_N$; b) equilibrium magnetic flux contours, red: last closed flux-surface, blue: first wall, green: simulation domain used in JOREK; c) Schematic showing the cross-section of the plasma along with the active and passive conducting structures of ITER used in the simulations. Note that the thick coils are discretized by several bands of thin structures in STARWALL. The acronyms used are VV: inner and outer shells of the vacuum vessel; PF: poloidal field coils; CS: central solenoid; D: divertor inboard rail; O: outer triangular support.}
\label{fig:equilibrium}
\end{figure}

For the setup used in this study, an elongated ITER plasma (pure deuterium) is considered  that is in free-boundary equilibrium, with a toroidal plasma current $I_p=14.6$MA, a uniform electron density of $n_e=10^{20}\mathrm{m}^{-3}$, an on-axis core electron temperature $T_{e,\mathrm{core}}\approx15$keV, central safety factor $q_0 \approx 1$ and an edge safety factor $q_\mathrm{edge}\approx 3.7$. The conducting structures (around the plasma) considered are the vacuum vessel (both the inner and outer shells), the poloidal field (PF) coils (PF1 to PF6), internal vertical stability coils (VS3), the central solenoid (CS1 to CS6), the outer triangular support (OTS) and the divertor inboard rail (DIR) \cite[]{Artola:2022}. Among these, the vacuum vessel, OTS and DIR are passive structures, the most important in the present context of vertical stability in ITER being the vacuum vessel with an electrical resistivity $\eta_\mathrm{wall}=0.8 \times 10^{-6} \mathrm{\Omega m}$ and a thickness of $6$cm (each shell). The equilibrium plasma profiles and magnetic flux contours are shown respectively in Fig.~\ref{fig:equil_Te_profile} and Fig.~\ref{fig:equil_flux_surfaces}, while the configuration of the conducting structures is shown in Fig.~\ref{fig:cond_structures}.

For the initial $1.5$ms in the simulation, the system is evolved from the initial static-equilibrium state resulting from the free-boundary Grad-Shafranov solver in JOREK to a state (that largely resembles the initial state) including the $\bm{E}\times \bm{B}$ flows that arise naturally.

At this time point, the various simulations can be seen as bifurcating into major disruptions (MD) and hot-VDEs. For the major disruptions (MD), an artificial thermal quench and current flattening is initiated immediately. For the hot-VDEs, first a current perturbation is applied to the internal vertical stability coils (VS3) over a duration of $~0.7$ms. This causes a loss of vertical stability and a subsequent vertical motion either upwards or downward depending on the sign of the imposed perturbation. For these hot-VDE cases, the artificial thermal quench and current flattening is initiated $50$ms after the plasma axis has drifted by $\Delta Z_\mathrm{a}=0.2$m from its initial equilibrium state. Such a timing corresponds to a detection of a hot-VDE by ITER control system within  a vertical plasma movement of $0.2$m and a subsequent $50$ms lag for the arrival of impurities injected by the ITER Disruption Mitigation System (DMS).

As mentioned earlier, at this time point, corresponding to $t=1.5$ms for MD and $t=t_{(Z_a=0.2m)}+50$ms for hot-VDEs, the plasma is artificially cooled via a large perpendicular thermal diffusivity over a duration of $0.5$ms so as to attain an on-axis electron temperature $T_{e,\mathrm{core}} \approx 20$eV. During this phase, the current density profile is simultaneously flattened artificially through the use of a large electrical hyperresistivity (different levels of current profile flattening were explored).  At the end of this phase, the artificial cooling is switched off, Ohmic heating is switched-on and a mixture of Neon ($5\times10^{21}$ or $1.5\times 10^{23}$ atoms) plus deuterium ($2\times10^{23}$ atoms) as well as an RE seed population ($0.1$A) is introduced within the whole computational domain with a spatially uniform-density.  This is intended to mimic an idealized massive material injection in the context of the ITER DMS.  The injected Neon and Deuterium corresponds to a $10\%$ assimilation of a $28\mathrm{mm}$ ITER DMS pellet and a density rise of $\Delta n_\mathrm{Ne}=4.51\times10^{18} \mathrm{m}^{-3}$ or $\Delta n_\mathrm{Ne}=1.35\times10^{20} \mathrm{m}^{-3}$ and $\Delta n_\mathrm{D}=1.81\times10^{20} \mathrm{m}^{-3}$ respectively. Note that full simulations (until RE termination) of MD (up) and hot-VDEs (up and down) are performed both with $\Delta n_\mathrm{Ne}=4.51\times10^{18} \mathrm{m}^{-3}$ and a much larger quantity $\Delta n_\mathrm{Ne}=1.35\times10^{20} \mathrm{m}^{-3}$ to explore the generation of RE beams at larger RE current.  After the injection, due to impurity radiation, the plasma tends to further cool down to lower temperatures competing with Ohmic heating. At the same time, due to the increased plasma resistivity at low-temperatures, the plasma current decays (current quench), along with the conversion of thermal current to RE current via the avalanche mechanism. During and after the thermal quench, the plasma becomes further vertically unstable and moves upwards/downwards. This leads to a continuous shrinking of the plasma column due to scraping-off at the wall, and the eventual loss of the whole plasma. 

Simulations are also performed with either 2\textsuperscript{nd} injection of Neon or 2\textsuperscript{nd} Deuterium injection via Neon-flushout for the MD and upward-hot-VDE cases. This is executed either midway during the RE avalanche (when the RE current is roughly half the peak plateau RE current) or once the RE current reaches its plateau phase. The Neon flushout is modeled via a density sink term (negative source). The list of simulations and their main differences are summarized in table.~1. The runs with the larger quantity of Neon are numbered with the suffix 'L' as shown in table.~1. The downward-hot-VDE case with $\Delta n_\mathrm{Ne}=4.51\times10^{18} \mathrm{m^{-3}}$ doesn't produce any noticeable RE beam and so it doesn't make sense to perform 2\textsuperscript{nd} Ne injection or Ne flushout for that case.

\begin{table}
  \begin{center}
\def~{\hphantom{0}}
  \begin{tabular}{lccccc} \hline
      \textbf{Run} & \textbf{Type}  & $ \mathbf{\Delta n_\mathrm{Ne} (10^{19} m^{-3})}$ &  $\mathbf{\Delta li}$ & \textbf{2\textsuperscript{nd} inj./flush} & $\approx \bm{I_{RE}^{max}}$ \textbf{(MA)} \\[3pt] \hline
       0   & MD Up (REs off) & $0.45$ & $0.28$ & - & -  \\
       1a   & MD Up ($3\%$ 1\textsuperscript{st} Ne) & $0.45$ & $0.28$ & - & 0  \\        
       1   & MD Up & $0.45$ & $0.28$ & - & 5.5  \\
       2  & MD Up & $0.45$ & $0.28$ & Ne-2\textsuperscript{nd}-P & 5.5   \\
       4   & MD Up & $0.45$ & $0.28$ & Ne-2\textsuperscript{nd}-M  & 5.3  \\
       5 & MD Up & $0.45$  & $0.28$ & D-2\textsuperscript{nd}+Ne-Flush-P & 5.5   \\
       6 & MD Up & $0.45$  & $0.28$ & D-2\textsuperscript{nd}+Ne-Flush-M  & 5.4  \\
       7   & MD Up & $0.45$  & $0.40$ & - & 5.1  \\
       8   & MD Up & $0.45$  & $0.03$ ($q < 2$) & - & 5.7  \\
       9   & Hot-VDE Up & $0.45$  & $0.36$ & - & 3.6  \\
       10   & Hot-VDE Up & $0.45$  & $0.36$ & Ne-2\textsuperscript{nd}-P & 3.6 \\
       11   & Hot-VDE Up & $0.45$  & $0.36$ & D-2\textsuperscript{nd}+Ne-Flush-P & 3.6   \\
       12   & Hot-VDE Dw & $0.45$  & $0.32$ & -  & 0.003 \\\hline  
       1L   & MD Up & $13.55$   & $0.28$ & - & 9.4  \\
       2L  & MD Up & $13.55$    & $0.28$ & Ne-2\textsuperscript{nd}-P & 9.4  \\
       4L   & MD Up & $13.55$    & $0.28$ & Ne-2\textsuperscript{nd}-M  & 7.9  \\
       5L & MD Up & $13.55$    & $0.28$ & D-2\textsuperscript{nd}+Ne-Flush-P & 9.4   \\
       6L & MD Up & $13.55$    & $0.28$ & D-2\textsuperscript{nd}+Ne-Flush-M  & 8.8  \\
       9L   & Hot-VDE Up & $13.55$    & $0.36$ & - & 9.0  \\
       10L   & Hot-VDE Up & $13.55$    & $0.36$ & Ne-2\textsuperscript{nd}-P & 9.0 \\
       11L   & Hot-VDE Up & $13.55$    & $0.36$ & D-2\textsuperscript{nd}+Ne-Flush-P & 9.0  \\
       12L   & Hot-VDE Dw & $13.55$    & $0.32$ & -  & 8.5 \\
       13L   & Hot-VDE Dw & $13.55$    & $0.32$ & -  Ne-2\textsuperscript{nd}-P & 8.5 \\
       14L   & Hot-VDE Dw & $13.55$    & $0.32$ & D-2\textsuperscript{nd}+Ne-Flush-P  & 8.5 \\\hline      
  \end{tabular}
  \caption{List of simulations including major disruptions (MD), up and downward hot-VDEs. The third column refers to the Neon density rise in the 1\textsuperscript{st} injection during which the Deuterium density rise is $\Delta n_\mathrm{D}=1.81\times 10^{20} \mathrm{m}^3$. The fourth column refers to the change in the plasma internal inductance $\Delta li$ during the TQ phase that indicates the extent of current profile flattening.  Terminology: \emph{Ne-2\textsuperscript{nd}-P} (2\textsuperscript{nd} injection of Neon in the early RE plateau phase); \emph{Ne-2\textsuperscript{nd}-M} (2\textsuperscript{nd} injection of Neon when about half the expected RE current is reached);  \emph{D-2\textsuperscript{nd}+Ne-Flush-P} (2\textsuperscript{nd} injection of Deuterium along with complete flushout of Neon starting in the early RE plateau phase);  \emph{D-2\textsuperscript{nd}+Ne-Flush-M} (2\textsuperscript{nd} injection of Deuterium along with complete flushout of Neon starting when about half the expected RE current is reached). Run names with the suffix 'L' indicate 30 times larger Neon 1\textsuperscript{st} injection quantity than those without the suffix. }
  \label{tab:kd}
  \end{center}
\end{table}

\subsection{Further details of the simulations}

 The artificial or pre-1\textsuperscript{st}-injection cooling of the plasma (e.g. from $t=1.5$ms to $t=2$ms in MD simulations) is achieved by the use of a large perpendicular thermal diffusivity. During this phase of the simulation, the parallel thermal diffusivity is increased simultaneously, such that there is no accumulation of thermal energy lost from the plasma core in the scrape-off layer (SOL) and the resistivity consequently remains high there. This ensures that the SOL is relatively cooler and the halo currents are minimised. The choice to minimize halo currents was made in order to maximize RE generation, since the current spread into the halo region could decrease the available electric field in the confined plasma. The plasma electrical resistivity ($\eta$) used is a function of temperature and $Z_\mathrm{eff}$, and is given by 
   $\eta_\mathrm{sp} = \eta_\mathrm{sp,c} \times Z_\mathrm{eff}$,
where $\eta_\mathrm{sp,c}$ is the classical Spitzer resistivity in the absence of impurities. In this respect, the dependence of Coulomb logarithm on temperature and density is not considered and instead a constant value ($\ln \Lambda =20$) has been used, that corresponds to a hot plasma at $T_e=15$KeV and $n_e=2\times10^{20}\mathrm{m^{-3}}$. The background plasma-fluid viscosity ($\mu=5.2\times 10^{-7}$kg-m/s) and parallel thermal diffusivity ($\kappa_\parallel = 1.54 \times 10^{29} \mathrm{m^{-1} s^{-1}}$) are chosen to be temperature independent, and there are no volumetric sources of thermal energy besides Ohmic heating and radiation losses. The SOL diffusivities are not treated differently. Standard fixed boundary conditions are used for the temperature, with the far SOL $T_e\approx 2$eV.

A spatially uniform RE seed current of $0.1$A has been used throughout and intends to represent the primary seed of REs that would have been generated during the thermal quench via the mechanisms of hot-tail, Compton scattering, Tritium decay and Dreicer \cite[]{Breizman:2019}. While a significant variation in the RE seed is possible in reality, the plateau RE current has been observed to have a very weak dependence on the RE seed \cite[]{Bandaru-model:2024}, justifying the simpler choice made here. To reduce computational costs for the long timescales simulated here, RE parallel transport is modeled via a large parallel diffusivity (to mimic the fast parallel advection). A value of $D_\mathrm{RE,||}=1.54\times 10^{6}\mathrm{m^2/s}$ has been used for the results presented in this work, along with a small perpendicular diffusivity of $D_\mathrm{RE,\perp}=4.6\times 10^{-2}\mathrm{m^2/s}$ for numerical stability. The choice for this value has been arrived through a sensitivity study that showed that much higher values would cause artificial perpendicular diffusion of REs in the later stages of the simulation. The perpendicular transport of REs occurs through the $\bm{E}\times \bm{B}$ advection of the RE fluid. 

 As mentioned earlier, for the sake of simplicity, Deuterium neutrals are not considered. This is justified by the fact that for the temperatures involved in this simulation, the neutral population is expected to be small and will have an insignificant effect on the final results. This has indeed been confirmed by comparing with a test simulation involving Deuterium neutrals. Also, we do not consider the parallel velocity of the background plasma in these simulations. Several test simulations showed that use of the customary fixed-boundary condition leads to large density gradients near the domain boundary leading to a significant loss of particles over the long timescale of the simulation. To avoid this, natural or Neumann condition have been used for the main-ion (D) density and impurity (Ne) density evolution equations, which ensures strict particle conservation in the domain. This is due to the choice of vanishing of flows normal to the domain boundary. Along with this, we use large values of particle diffusivity to maintain a nearly uniform density distribution. Values of $D_\mathrm{\perp}=15.4\mathrm{m^2/s}$ and $D_\mathrm{\parallel}=1.54\times10^{5}\mathrm{m^2/s}$ respectively have been used for both the main-ions and impurities.

A polar grid with radial and poloidal grid resolution $N_r \times N_\theta = 101\times 128$ bi-cubic Bezier finite-elements is used with poloidal grid clustering in the region covered by the plasma during its vertical trajectory. We now to turn to the discussion and analysis of the results obtained from the simulations.

\section{Results and discussion}

\subsection{General characteristics and the effect of RE beam formation}
\begin{figure}
\centering
\includegraphics[trim = 0mm 0mm 0mm 0mm, clip, width=0.6\textwidth]{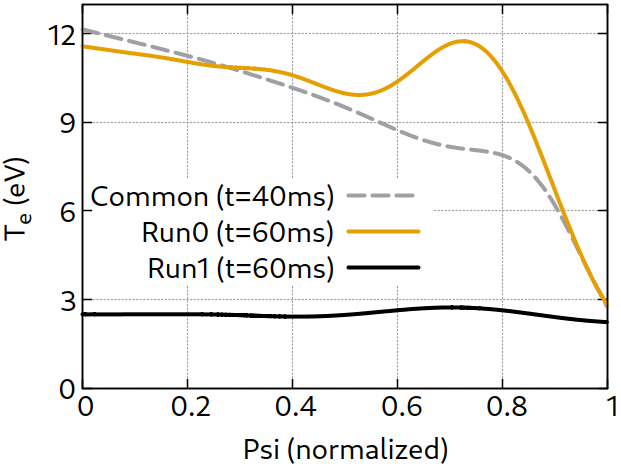}
\caption{Electron temperature profiles before and after RE formation for Run1 compared to the corresponding case wherein REs are switched-off (Run0).}
\label{Te_profiles_run0_run1}
\end{figure}

Firstly we look into the general physical behaviour in our simulations, wherein the focus is on how RE formation modifies the overall evolution of a disruption. As mentioned earlier, the artificial thermal quench executed over a time of $\approx 0.5$ms causes the on-axis temperature to drop to $T_{e,\mathrm{core}} \approx 20$eV. After the 1\textsuperscript{st} injection of Deuterium and Neon, impurity radiation dominates over Ohmic heating causing a further cooling of the plasma to settle at roughly $T_{e,\mathrm{core}} \approx 10$eV. Subsequently RE formation causes cooling of the plasma to continue to further lower temperatures due to the decrease in thermal plasma current and associated Ohmic heating, such that the effective ion-charge is around unity even in the presence of impurities. Figure.~\ref{Te_profiles_run0_run1} compares the electron temperature profile after the full RE beam formation for a major disruption case (Run1) to that when RE avalanche is switched-off (Run0). Additional cooling due to RE beam formation is evident. Moreover, the formation of RE beam also results in a decrease in plasma internal inductance $l_i$ due to the general tendency of a peaked RE current profile formation. This can in principle affect the kink stability of the current beam, e.g. due to a relatively lower central safety factor.

\begin{figure}
\centering
\includegraphics[trim = 0mm 0mm 0mm 0mm, clip, width=1.0\textwidth]{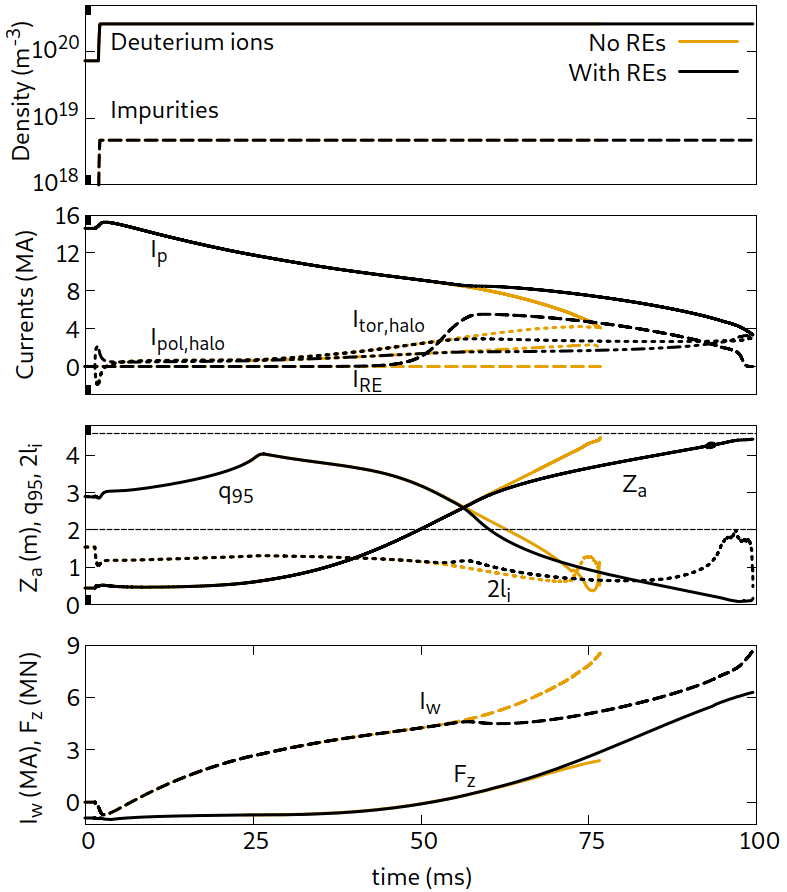}
\caption{Comparison of the evolution of 0D parameters during the MD cases with REs (Run1, black) and with REs switched-off (Run0, orange). Nomenclature: $I_p$ is the total plasma current, $I_\mathrm{pol,halo}$ and $I_\mathrm{tor,halo}$ are the poloidal and toroidal halo currents respectively, $I_\mathrm{RE}$ is the RE current, $q_\mathrm{95}$ is the safety factor at $\psi_N=0.95$, $Z_a$ is the vertical position of the plasma magnetic axis, $l_i$ the plasma internal inductance, $I_w$ the wall current and $F_z$ the total vertical force on the vacuum vessel.}
\label{fig:R0_vs_R1_all}
\end{figure}

\begin{figure}
\centering
\includegraphics[trim = 0mm 0mm 0mm 0mm, clip, width=0.6\textwidth]{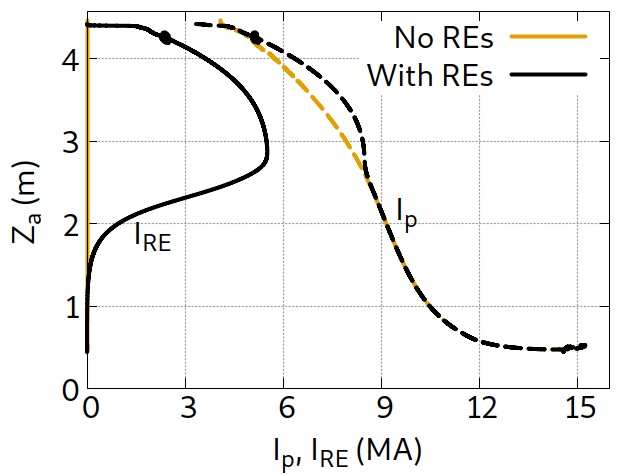}
\caption{Total plasma current ($I_p$) and RE current ($I_\mathrm{RE}$) versus the vertical position of the plasma magnetic axis ($Z_\mathrm{a}$) for the MD cases with REs (Run1) and without REs (Run0).}
\label{fig:R0_vs_R1_Ip_Z}
\end{figure}

\begin{figure}
\centering
\includegraphics[trim = 0mm 0mm 0mm 0mm, clip, width=0.65\textwidth]{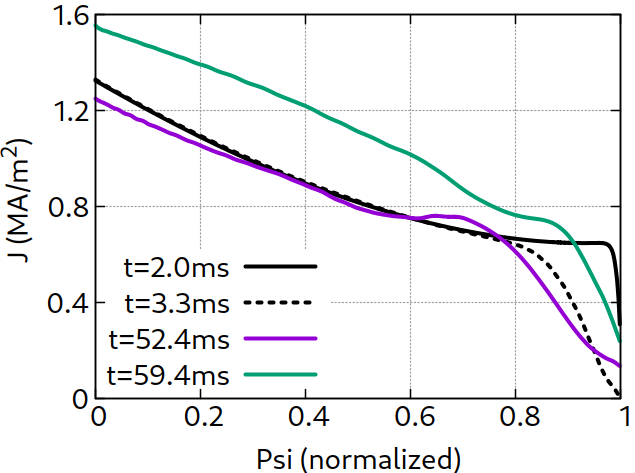}
\caption{Profiles of plasma current density for Run1 (MD) at various time points from the end of thermal-quench till RE plateau formation. Bold black: immediately after artificial-TQ and current flattening; Dotted black: slightly after the end of 1\textsuperscript{st} injection of Ne+D; Magenta: just before any significant RE beam formation; Green: after full RE beam formation.}
\label{jprofiles_run1}
\end{figure}

Figure~\ref{fig:R0_vs_R1_all} shows the subsequent effects of RE formation as compared to a situation when REs are switched-off. It can be observed that the primary effect of RE beam formation is a decrease in the current decay rate post the beam-formation. This is due to the fact that REs are not subjected to a decay on the resistive timescale. This also results in a relative slowing down of the plasma vertical motion towards the wall, since the current decay rate is a significant driving cause for the plasma vertical motion in ITER due to the relatively long wall time $\sim 500$ms. The slower plasma current decay rate with REs  results in the corresponding slower increase in the toroidal wall current $I_w$ (see the last panel of Fig.~\ref{fig:R0_vs_R1_all}). However the total plasma current is roughly the same in both cases at the time when the current beam collapses into the wall. This can be clearly seen in Fig.~\ref{fig:R0_vs_R1_Ip_Z}, wherein the currents at any given vertical  location of the plasma axis are plotted. Therefore RE beam formation makes the plasma take longer to reach the wall (an additional $\approx 23$ms in this specific case), but with roughly the same total plasma current. 

Figure.~\ref{jprofiles_run1} shows the profiles of current density for Run1 at different time points until the RE plateau formation. The J-profile after avalanche is a result of both the avalanche as well as the minor-radial diffusion of the parallel electric field during that phase. One can observe that the J-profile after full RE beam formation (green curve) is centrally-peaked. Note that the magnitude of $J$ increases due to the plasma shrinking already during this phase. In an axisymmetric situation, one relevant/important force on the vacuum vessel is the vertical force $F_z$. However it is important to note that the force on the vacuum vessel peaks only after the change in magnetic field diffuses into the vessel material, which occurs over $\tau_w = L/R \sim 500$ms timescale \cite[]{Pustovitov:2017, Clauser:2019, Artola:2022}. The simulations performed here, though relatively long-timescale, span until the final collapse of the current beam which is $\sim 100-150$ms. This does not allow for a peaking of the wall force, and hence the force values seen shouldn't be interpreted as the maximum forces predicted.

\begin{figure}
\subfigure[]{
\centering
\includegraphics[trim = 0.6mm 0mm 0mm 0mm, clip, width=0.5\textwidth]{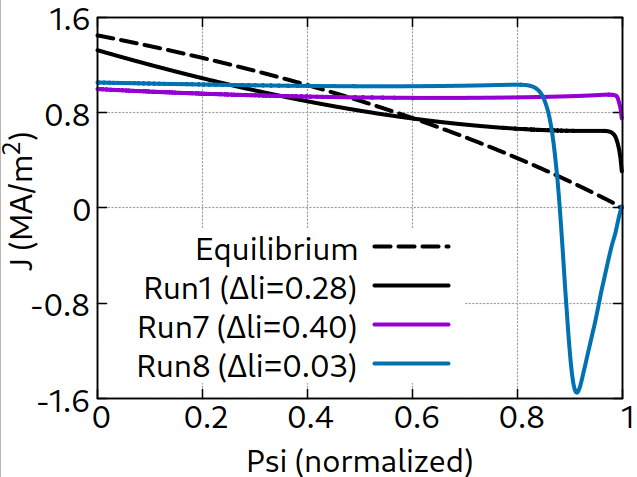}
\label{fig:Jprofiles_diff_flatt}
}
\subfigure[]{
\centering
\includegraphics[trim = 0mm 0mm 0mm 0mm, clip, width=0.5\textwidth]{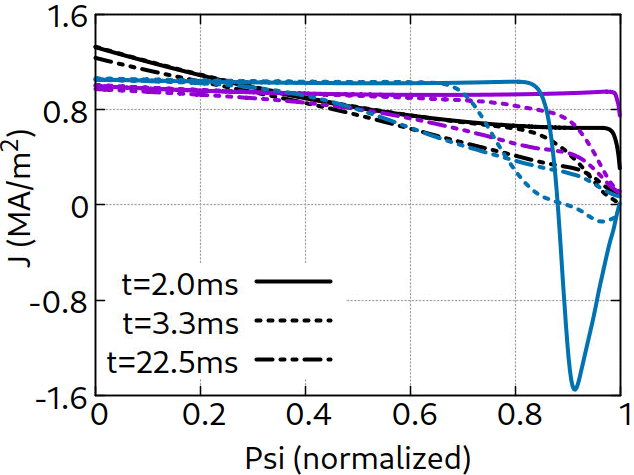}
\label{fig:Jprofile_relazation}
}
\caption{a) Profiles of plasma current density with different level of flattening immediately after the artificial thermal-quench ($t=2$ms) as a function of normalized poloidal flux; b) Relaxation of post thermal-quench flattened profiles of plasma current density as a function of normalized poloidal flux. Black: Run1, Magenta: Run7, Blue: Run8.}
\end{figure}

\subsection{Impact of the post thermal quench current density profile}

It is well-established that the aspect of magnetic helicity conservation during the fast thermal quench (TQ) phase of a disruption, tends to flatten the current density profile relative to pre-TQ profile \cite[]{Biskamp:1993}. In this regard, we explore the effect of the immediate post-TQ current density profile on further evolution of the disruption. What might be the most realistic current flattening scenario for a real disruption would likely depend on the details of MHD activity and magnetic stochastization before and during the thermal quench. For lack of known constraints on the post-TQ current profile, a few different current profile flattening scenarios during the artificial thermal quench phase have been chosen in the context of major disruptions. They include the cases of significant flattening (Run1), full-flattening (Run7) and full-flattening but only within the region enclosed by the $q=2$ flux surface (Run8). The corresponding post-TQ current density profiles are shown in Fig.~\ref{fig:Jprofiles_diff_flatt}, wherein they are characterized also by the change in internal inductance $\Delta li$ after the flattening. 

\begin{figure}
\centering
\includegraphics[trim = 0mm 0mm 0mm 0mm, clip, width=1.0\textwidth]{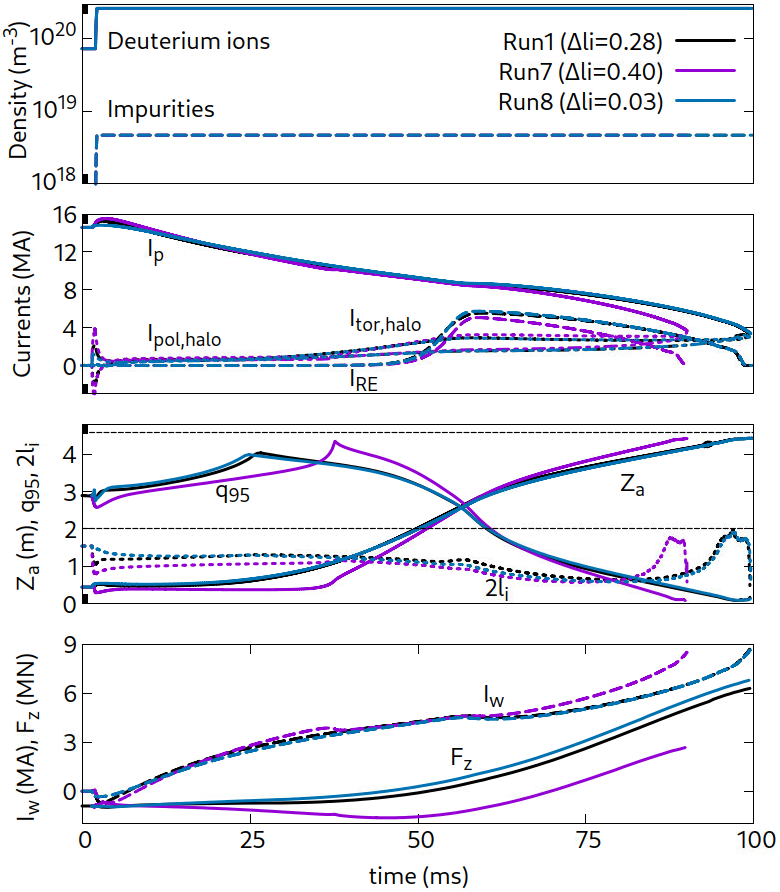}
\caption{Comparison of the evolution of 0D parameters for the major disruption (MD) cases with different current profile flattening during the artificial thermal quench. Due to fast current profile relaxation post flattening, the difference in evolution between Run1 and Run8 is negligible, while Run7 evolves differently due to different wall current distribution.}
\label{fig:R1_vs_R7_vs_R8_all}
\end{figure}

\begin{figure}
\centering
\includegraphics[trim = 0mm 0mm 0mm 0.5mm, clip, width=0.65\textwidth]{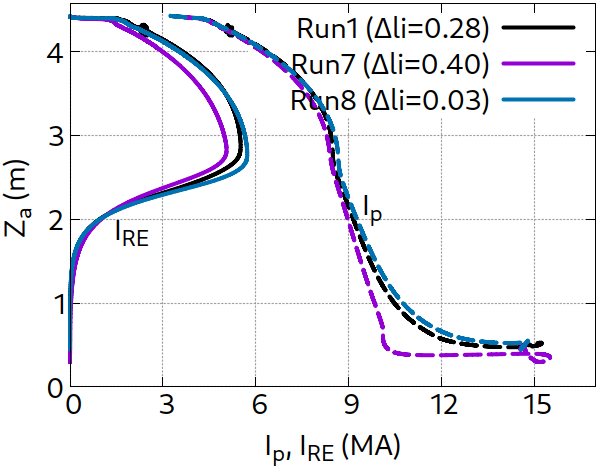}
\caption{Total plasma current and RE current versus the plasma axis vertical position for the MD cases with different current profile flattening during the artificial thermal quench.}
\label{fig:R1_vs_R7_vs_R8_Ip_Z}
\end{figure}

The differences in the overall evolution in these cases is shown in Fig.~\ref{fig:R1_vs_R7_vs_R8_all} and Fig.~\ref{fig:R1_vs_R7_vs_R8_Ip_Z}. It can be seen that, except for a short time window of $\sim10$ms post TQ, there is negligible difference in the evolution of Run1 (black) and Run8 (blue). This can be attributed to the relaxation (via resistive diffusion) of the current density profiles after the flattening. Both the profiles develop large near-edge gradients in the current density immediately after flattening (Fig.~\ref{fig:Jprofiles_diff_flatt}). However, after the artificial TQ and the subsequent 1\textsuperscript{st} injection of Neon, the plasma resistivity is rather large and this ensures that the near edge gradients are smoothed-out within $\sim20$ms. This is shown in Fig.~\ref{fig:Jprofile_relazation}, where in fact by $t\approx22.5$ms, the core J-profiles  of all the cases are very similar, and in particular the profiles of Run1 and Run8 match very closely.

In spite of this effect, interestingly, the evolution of the case with fully flat J-profile (Run7) shows significant differences over the course of the simulation. This is due to a different distribution of current density on the plasma, which leads to a different wall current distribution when the plasma current decays. Such current density distributions delay the vertical movement (it even moves downwards during the initial phase after reversing its direction to upwards).  One can see Run7 to be a case of being near the threshold between upward and downward motion. In summary, while the precise post-TQ J-profile can have an effect on the further evolution of the disruption, fast radial resistive diffusion of current tends to suppress the differences to a great extent.
We now turn to the effects of 2\textsuperscript{nd} injection of massive material into the plasma.

\subsection{Neon 2\textsuperscript{nd} injection and Neon flushout}

\subsubsection{Neon secondary injection}
\begin{figure}
\centering
\includegraphics[trim = 1mm 15mm 0mm 0mm, clip, width=0.87\textwidth]{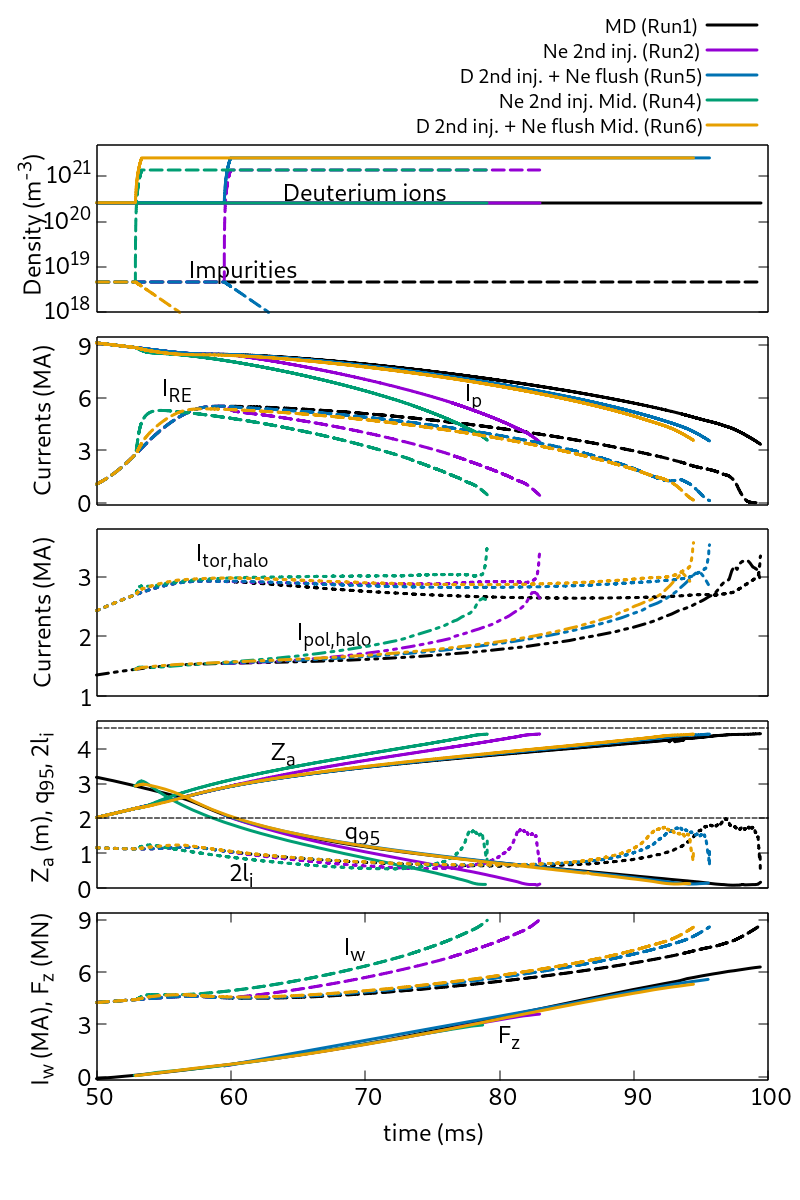}
\caption{Comparison of the evolution of 0D parameters for the MD cases with Ne 2\textsuperscript{nd} injection and Ne-flushout with D 2\textsuperscript{nd} injection both starting at the plateau phase and midway during the RE avalanche}
\label{fig:MD_2ndinj_flush_all}
\end{figure}

Second injection of massive quantities of Neon through shattered pellets (SPI) onto an existing RE beam is a strategy to ensure a faster decay of the RE beam current. In our simulations, for the Neon 2\textsuperscript{nd} injection cases, a total of $1.5\times10^{24}$ Neon atoms are introduced uniformly into the computational domain (assuming a $100\%$ assimilation and corresponds to a Neon density rise $\Delta n_\mathrm{Ne}=1.35\times10^{21} \mathrm{m^{-3}}$). This is done via an impurity particle source activated over a relatively short duration. The injected Ne quantity amounts to a factor of $300$ times the existing Ne atoms from the 1\textsuperscript{st} injection, and about $5.4$ times the total existing D atoms in the plasma (see Fig.~\ref{fig:MD_2ndinj_flush_all} topmost panel). The relatively large quantity of Neon for 2\textsuperscript{nd} injection has been chosen intentionally to assess whether any beneficial effects could be expected at all. For the MD cases (branched from Run1), 2\textsuperscript{nd} Ne injection is made both at the beginning of the RE plateau phase (Run2) and at midway during RE avalanche i.e. when the RE current is half the expected peak value (Run4). For the upward VDE case (branched from Run9), 2\textsuperscript{nd} Ne injection is made only at the beginning of the RE plateau phase (Run10). It must be noted that before Ne 2\textsuperscript{nd} injection is made, the plasma is already cold ($<10$eV) and so only the lower charge-states of Ne exist, meaning that $Z_\mathrm{eff}\approx1$. So additional Ne injection does not alter $Z_\mathrm{eff}$ much and therefore the plasma electrical resistivity $\eta$ doesn't change significantly.

\begin{figure}
 \centering
\subfigure[]{
\centering
\includegraphics[trim = 0mm 0mm 0mm 0mm, clip, width=0.65\textwidth]{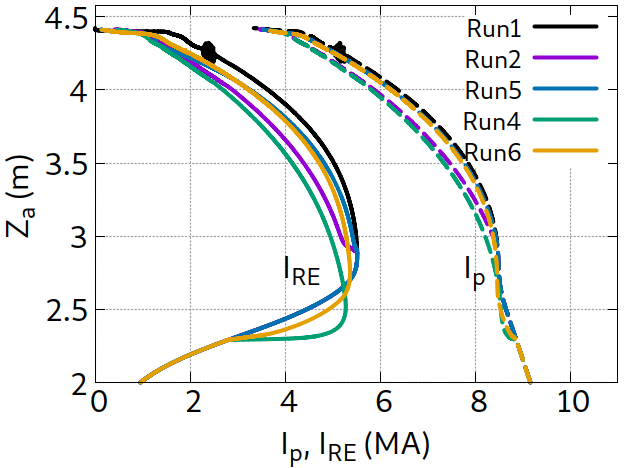}
\label{fig:MD_2ndinj_flush_Za_Ip}
}

\subfigure[]{
\centering
\includegraphics[trim = 0mm 0mm 0mm 0mm, clip, width=0.65\textwidth]{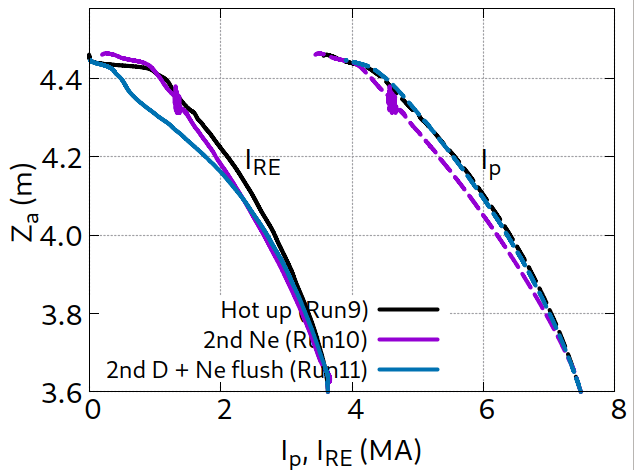}
\label{fig:Up_2ndinj_flush_Za_Ip}
}
\caption{Total plasma current and RE current versus the plasma axis vertical position for a) MD cases with Ne 2\textsuperscript{nd} injection and Ne-flushout with D 2\textsuperscript{nd} injection starting both at the plateau phase and midway during the RE avalanche; b) upward-VDE cases with Ne 2\textsuperscript{nd} injection and Ne-flushout with D 2\textsuperscript{nd} injection starting at the RE plateau phase.}
\label{fig:2ndinj_flush_Za_Ip}
\end{figure}

Comparison of the effect of 2\textsuperscript{nd} Ne injection in MD cases (Run1, Run2, Run4) is shown in Fig.~\ref{fig:MD_2ndinj_flush_all} and Fig.~\ref{fig:MD_2ndinj_flush_Za_Ip}.
For the case with plateau injection (Run2), introduction of massive quantities of Neon causes a significantly faster decay of the RE beam current as can be seen from Fig.~\ref{fig:MD_2ndinj_flush_all} due to the increase in the effective critical electric field. This correspondingly decreases the timescale of the vertical motion and so the plasma dumps into the wall much faster too, as compared to the case without 2\textsuperscript{nd} injection. In this case, the net effect is that the RE current just before the final loss of the confined plasma region is lower $\sim1$MA versus $1.5$MA for the case without 2\textsuperscript{nd} injection as can be seen from Fig.~\ref{fig:MD_2ndinj_flush_Za_Ip}. The effect is similar for the case of plateau Ne 2\textsuperscript{nd} injection onto the upward VDE (Run10). However, in the upward VDE case, the RE current before final collapse is only marginally lower than the corresponding case without 2\textsuperscript{nd} Ne injection (see Fig.~\ref{fig:Up_2ndinj_flush_Za_Ip}). This can be attributed to a slightly lower stabilization factor $f$ \cite[]{Leuer:1989} in the case of upward VDE, which determines the growth rate of the vertical motion.

When Ne is introduced halfway during the avalanche growth (Run4), the immediate effect is quite different in the sense that the RE avalanche growth rate increases dramatically. This is due to the fact that there is enough electric field left and the number of electrons available for avalanche (which also includes bound electrons) increases drastically. That is, the effect of additional bound electrons from the added Ne that aid as targets dominates the effect of corresponding higher effective critical field $E_\mathrm{c,eff}$. The peak RE current obtained is however the same. It must be noted that propensity to avalanche (due to larger Neon quantity and thereby more bound electrons) doesn't always imply that the final RE beam current should be higher. Parallel electric field falls due to RE avalanche and resistive current decay, and so the avalanche growth stalls once there isn't enough electric field left. Further from the RE plateau downstream, the behaviour is similar to the other 2\textsuperscript{nd} Ne injection cases, i.e. faster RE current decay and vertical motion.  

 In summary, Neon 2\textsuperscript{nd} injection in general causes faster RE beam decay and correspondingly faster vertical motion to the wall.  Earlier injection (during avalanche) enhances the avalanche rate but the subsequent evolution is similar to the plateau injection. 

\begin{figure}
\centering
\includegraphics[trim = 0mm 0mm 0mm 0mm, clip, width=1.0\textwidth]{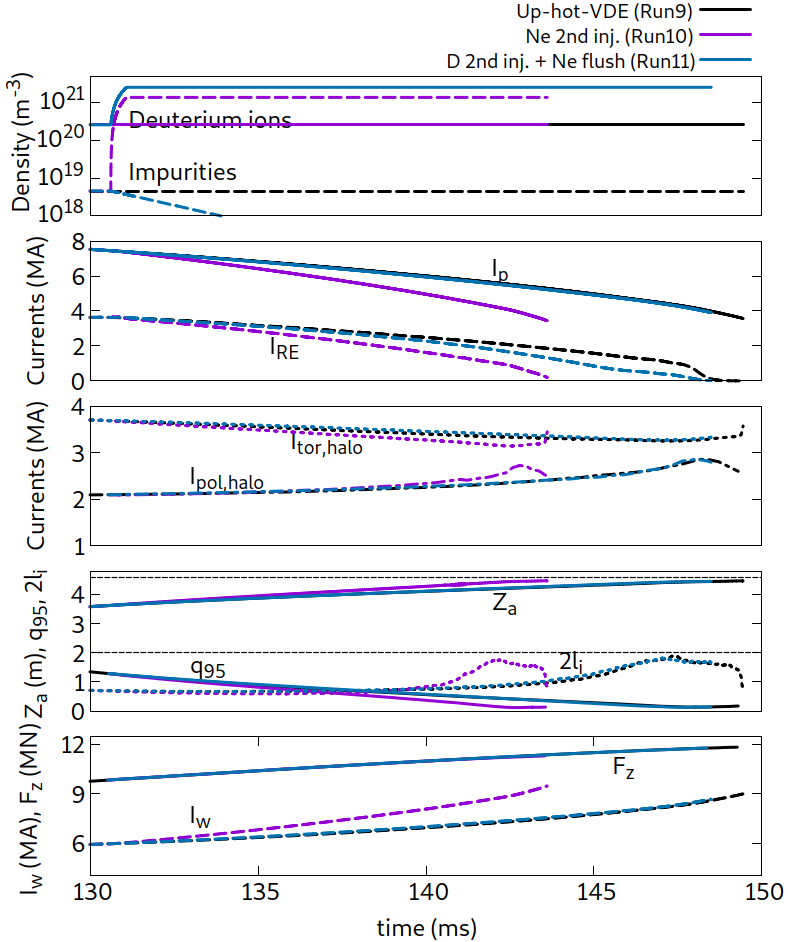}
\caption{Comparison of the evolution of 0D parameters for the upward-VDE cases with Ne 2\textsuperscript{nd} injection and Ne-flushout with D 2\textsuperscript{nd} injection.}
\label{fig:Up_2ndinj_flush_all}
\end{figure}

\subsubsection{Neon flushout induced by Deuterium secondary injection}
 In the recent years, impurity flushout through massive 2\textsuperscript{nd} injection of Deuterium has been found to show some promise in obtaining benign RE beam terminations (= no wall damage due to REs) in experiments at DIII-D and JET \cite[]{Reux:2021, Paz_Soldan:2021}. It appears that a benign termination after Ne flushout is caused not by a single or unique mechanism.  Some known pathways to such a benign termination post Ne flushout are hollow current-density profile formation that are inherently MHD unstable and a drop in $q_{95}<2$ but with a monotonous J-profile etc. All these pathways are presumed to result in a violent MHD activity that distributes the REs relatively broadly over the wall. Importantly, the experimental observation seems to be rather robust. In experiments, after Deuterium 2\textsuperscript{nd} injection, impurity flushout from the plasma occurs due to neutralization and subsequent outward transport of the neutral impurity atoms.  In our simulations, we model Ne flushout in a simplified way via an artificial Neon sink (negative source) to ensure an exponential loss of impurities on a $10$ms timescale (only about $1\%$ of impurities remain in the plasma after $10$ms). The timescale of impurity flushout in DIII-D and JET was observed to be $\sim10$ms. While there is no established consensus on what would be the expected Ne flushout timescale in ITER, it can be expected to be higher due to the relatively larger spatial scales. Nevertheless, our choice of the timescale $10$ms is driven by the fact that the total time to RE beam loss itself is at most $50$ms, and so a flushout on a much slower timescale wouldn't have a significant effect. A total of $2.5\times10^{24}$ Deuterium atoms are injected uniformly in the domain, which amounts to about $9$ times the existing D atoms in the plasma.

 In the present simulations, the effect of D 2\textsuperscript{nd} injection + Ne flushout is shown in Figs.~\ref{fig:MD_2ndinj_flush_all}-\ref{fig:2ndinj_flush_Za_Ip}, for both the MD cases and upward hot-VDE (Run5, Run6 and Run11). In the MD cases (Run5 and Run6) the impact is not as significant as Ne 2\textsuperscript{nd} injection and the results is a marginally faster RE current decay and corresponding vertical motion (see Fig.~\ref{fig:MD_2ndinj_flush_all}). In spite of a loss of Ne impurities, we see a faster RE current decay. This is due to the fact that the pre-existing quantity of Ne is relatively not large and the RE current decay is mainly dominated by the scrape-off rather than due to impurities. Hence the loss of impurities is offset by the massive quantity of D atoms injected. And so the net effect is a slightly faster decay of RE current. It must be noted that in many experiments a slow rise in RE current is observed post Ne flushout  (e.g. shot:95135 of JET, \cite{Reux:2021}). Such an RE current rise can be attributed to the fact that in many such RE experiments, there is no scrape-off effect since the plasma is vertically stable (nearly-circular limiter configuration).
 
 In summary, in comparison with Neon 2\textsuperscript{nd} injection, Neon flushout has a qualitatively similar but relatively weaker effect on the overall VDE. However, unlike in many existing experiments with circular plasmas, Neon flushout in elongated plasmas can in fact lead to a current decay rather than a current rise, mediated by the scrape-off effect.

 \subsubsection{Estimates for RE energy loss after Neon 2\textsuperscript{nd} injection and flushout}
 
 In the context of massive material injections, it is important to assess how much net energy will have been extracted by REs by the time they dump into the first wall. While REs gain energy via acceleration by the parallel electric field, they lose energy via the channels of synchrotron radiation, Bremsstrahlung and collisional drag.  The synchrotron loss (per particle) is strongly dependant on the energy and pitch angle of REs, and scales (for small pitch angles) as $\propto B^2 \theta^2 \gamma^2$, where $B$ is the magnetic field, $\theta$ the RE pitch angle and $\gamma$ is the relativistic factor. On the other hand, the Bremsstrahlung and collisional drag losses per particle scale as $\propto  n_e \gamma \left(Z_\mathrm{eff} + 1\right)$ and $\propto n_e \left(Z_\mathrm{eff} + 1 + \gamma \right) \gamma^{-1} $ respectively, again assuming a small pitch angle and high energy of REs.  Clearly, as mentioned earlier, both the pitch angle as well as energy of the REs are not tracked in the RE fluid model used in the present simulations (note that future work including REs in a PIC treatment in JOREK will improve the modelling fidelity). Nevertheless, it is still possible to make indirect rough estimates of the RE energy losses via the effective critical electric field $E_\mathrm{c,eff}$ that is computed within the RE fluid model while evaluating the avalanche source. That an estimate can be made using $E_\mathrm{c,eff}$ would be evident from the fact that the analytical model for $E_\mathrm{c,eff}$ used in our RE fluid model as derived by \cite{Hesslow:2018-PPCF} already includes the effects of synchrotron radiation, Bremsstrahlung and collisional loss. 

\begin{figure}
\centering
\includegraphics[trim = 0mm 0mm 0mm 0mm, clip, width=0.65\textwidth]{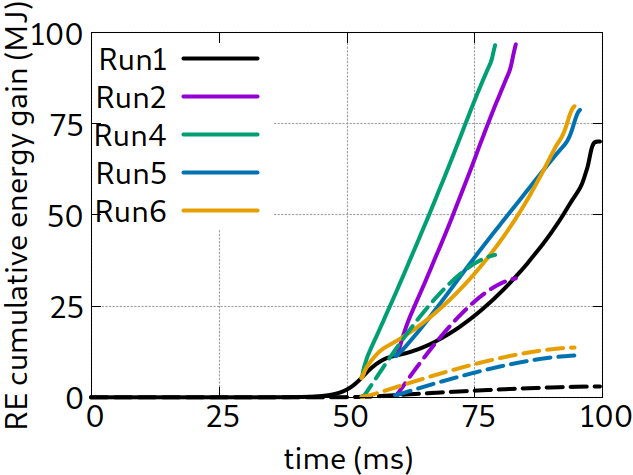}
\caption{Cumulative poloidal magnetic energy that is channeled to REs $\int J_{RE,\parallel} E_{\parallel} dV dt$ (bold lines) and the part $\int J_{RE,\parallel} E_\mathrm{c, eff} dV dt$ that is dissipated by REs via collisions and Bremsstrahlung (dashed lines), in different injection scenarios. Difference between the bold and dashed curves indicates the energy content of the REs at any given time.}
\label{fig:RE_energy}
\end{figure}

We now proceed to make RE energy loss estimates as follows. The parallel momentum balance for REs can be written as
\begin{align*}
    \frac{d}{dt} \left(\gamma m_{e0} v_{||} n_r \right) &= -en_r E_{||} - F_\mathrm{drag}
\end{align*}
Multiplying the above equation with $v_{||}$, using $F_\mathrm{drag}=-e n_r E_c^\mathrm{eff}$ and integrating over the computational volume, one can rewrite the above equation as
\begin{align*}
    \int \frac{d}{dt} \left(\frac{1}{2}\gamma m_{e0} v_{||}^2 n_r \right) dV dt &= \int J_{RE,||} \left(E_{||}-E_c^\mathrm{eff}\right) dV dt
\end{align*}
So the total poloidal magnetic energy channelled to REs is $\int J_{RE,||} E_{||} dV dt$ (which is the total work done by the electric field on the REs), out of which $\int J_{RE,||} \left(E_{||}-E_c^\mathrm{eff}\right) dV dt$ goes towards their kinetic energy and $\int J_{RE,||} E_c^\mathrm{eff} dV dt$ is the energy needed to sustain the RE current. At the end, this total energy channeled to REs gets lost by collisions, Bremsstrahlung, Synchrotron and deposition on PFCs via scrape-off and final-loss.

The total energy to REs in different injection scenarios is plotted in Fig.~\ref{fig:RE_energy} for the MD cases. Expectedly, both the cumulative energy channeled to REs (by the parallel electric field) as well as the cumulative energy dissipated by them increase over time, with a remnant net energy contained with the REs by the time of final termination. However, it can be observed that in all the cases shown, nearly $70-100$ MJ of the total poloidal magnetic energy $\sim 430$MJ is extracted by the REs by the time of final termination (the difference between the bold lines and dashed lines in Fig.~\ref{fig:RE_energy}). This can be attributed to the fact that while 2\textsuperscript{nd} injection causes an increase in RE energy dissipation, at the same time it also increases the total energy extracted by the REs from the poloidal magnetic field. Therefore it turns out that the net effect of 2\textsuperscript{nd} Ne injection and Ne flushout therefore is rather marginal, when it comes to the energy content of the REs by the time of final termination. Our observation here is qualitatively similar to the conclusion arrived by earlier work using DINA \cite[]{DINA-report:2016}, though there are several differences between both the models (e.g. DINA's model did not include the effects of partially-ionized impurities). Similar to the enhancement of total RE beam current formed after Ne 1\textsuperscript{st} injection (due to additional target electrons), sustained avalanche post Ne 2\textsuperscript{nd} injection (Run2, Run4) or D 2\textsuperscript{nd} injection/Ne flushout (Run5, Run6) ensures a slightly larger energy extracted by the REs. 

\subsection{Major disruptions versus hot-VDEs}

\begin{figure}
\centering
\subfigure[]{
\centering
\includegraphics[trim = 0mm 0mm 0mm 0mm, clip, width=0.31\textwidth]{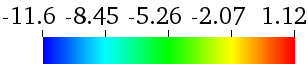}
}

\subfigure[]{
\centering
\includegraphics[trim = 0mm 0mm 0mm 0mm, clip, width=0.31\textwidth]{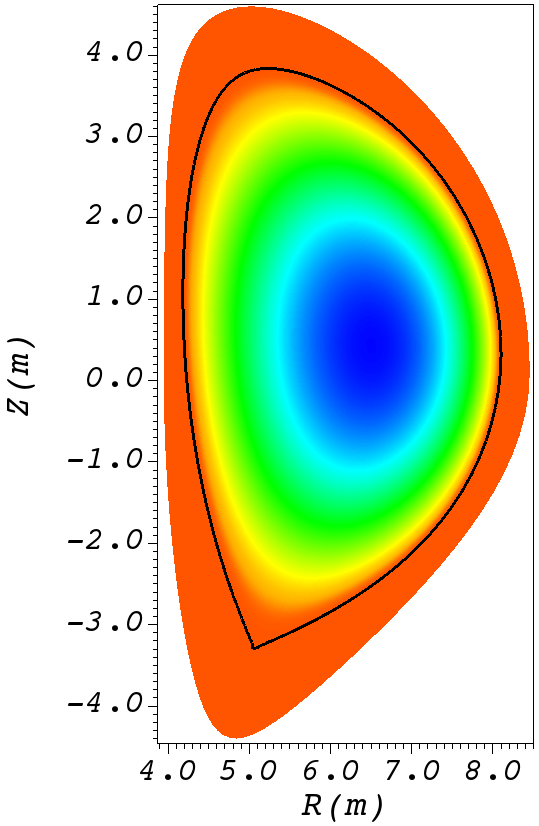}
}
\subfigure[]{
\centering
\includegraphics[trim = 0mm 0mm 0mm 0mm, clip, width=0.305\textwidth]{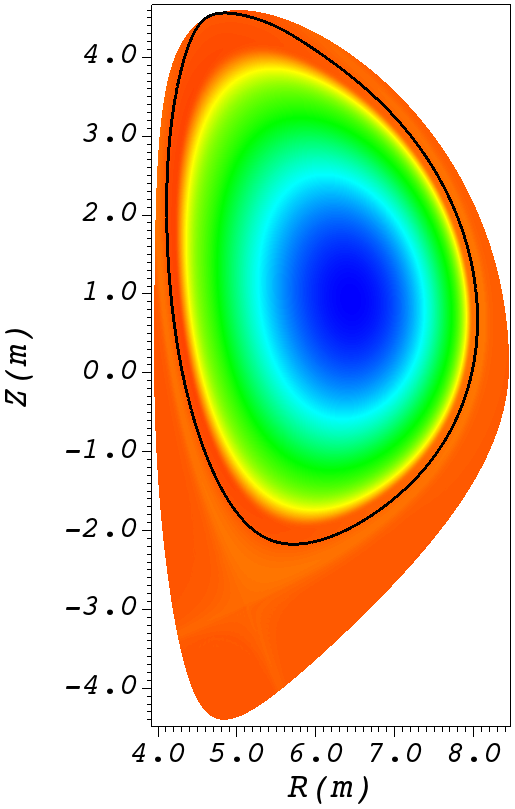}
}
\subfigure[]{
\centering
\includegraphics[trim = 0mm 0mm 0mm 0mm, clip, width=0.31\textwidth]{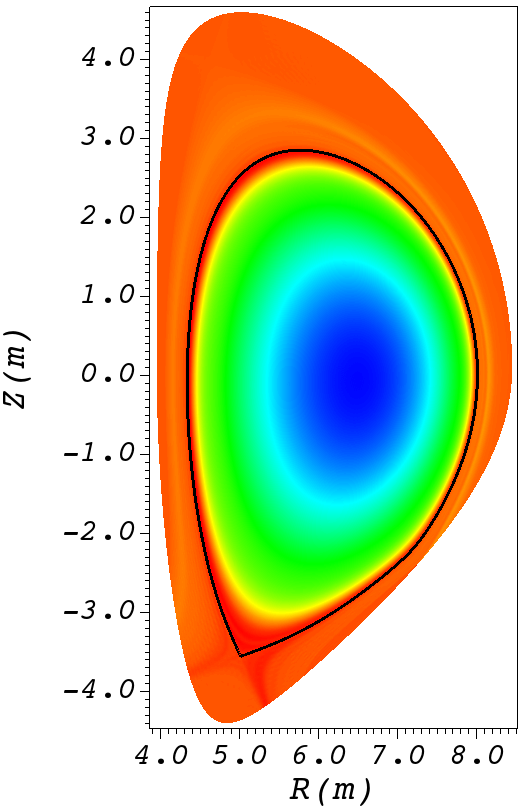}
}
\caption{Normalized toroidal current density distribution just before the initiation of artificial thermal quench for the a) MD (Run7), b) Up hot-VDE (Run9), and c) Down hot-VDE (Run12) cases.}
\label{fig:plasmaposition_md_up_down}
\end{figure}

\begin{figure}
\centering
\includegraphics[trim = 0mm 0mm 0mm 0mm, clip, width=0.65\textwidth]{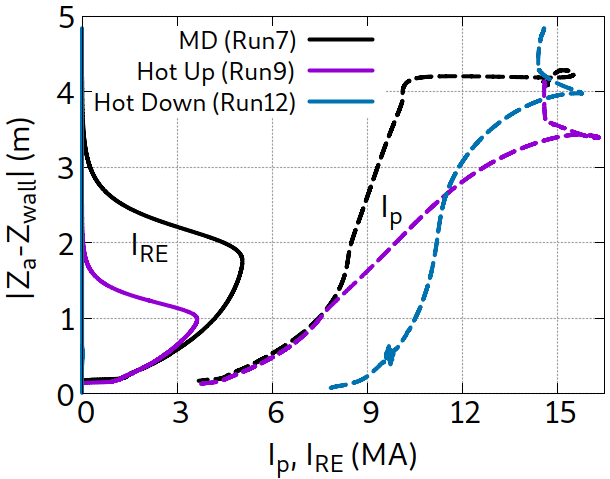}
\caption{Total plasma current and RE current versus the vertical distance of the plasma magnetic axis from the wall-touching-point, comparing the baseline cases (without 2\textsuperscript{nd} injections) of major disruption (MD) (Run7) with that of the hot-VDE up (Run9) and hot-VDE down (Run12).}
\label{fig:md_up_down_Ip_Z}
\end{figure}

In this section, we look into the main differences between the major disruption (Run7) and hot-VDE baseline cases (Run9 and Run12), all without any additional 2\textsuperscript{nd} injection of material. In the hot-VDE cases, effectively the artificial thermal quench gets initiated about $\sim 75$ms later than that in the MD case. During this time, due to the initially applied current perturbation in the internal vertical stability coils (VS3), the plasma in the hot-VDE cases drifts significantly (and reduces in size) already before the TQ is initiated. This is evident from the vertical parts of the dashed blue and magenta curves at nearly constant $I_p$ in Fig.~\ref{fig:md_up_down_Ip_Z}, wherein the axis distance from the wall is plotted versus the total and RE currents. The relative plasma positions just before the TQ is shown in Fig.~\ref{fig:plasmaposition_md_up_down} for all the 3 cases.

In the case of the upward hot-VDE (Run9), the overall evolution of the disruption is in general qualitatively similar to the MD case (Run7). There are however certain differences, that arise due to the delay in TQ initiation in the hot-VDE. A relatively larger $I_p$-spike is observed in Run9 as compared to Run7 due to the slightly higher flattening of the current profile during the artificial thermal-quench. Furthermore, due to the relatively smaller cross-section by the time of 1\textsuperscript{st} injection, the total assimilated Ne within the LCFS is smaller in this case (since the total number of injected atoms remains the same). As a consequence, the avalanche-magnifying effect of partially-ionized impurities is reduced and therefore results in a relatively smaller plateau RE current of $\sim3.7$MA versus $\sim5$MA in the case of MD. However, the later formation of RE current in Run9 eventually leads to a similar RE current at the time of final collapse ($\sim1$MA) as is evident from Fig.~\ref{fig:md_up_down_Ip_Z}.

\begin{figure}
\centering
\includegraphics[trim = 0mm 0mm 0mm 0mm, clip, width=1.0\textwidth]{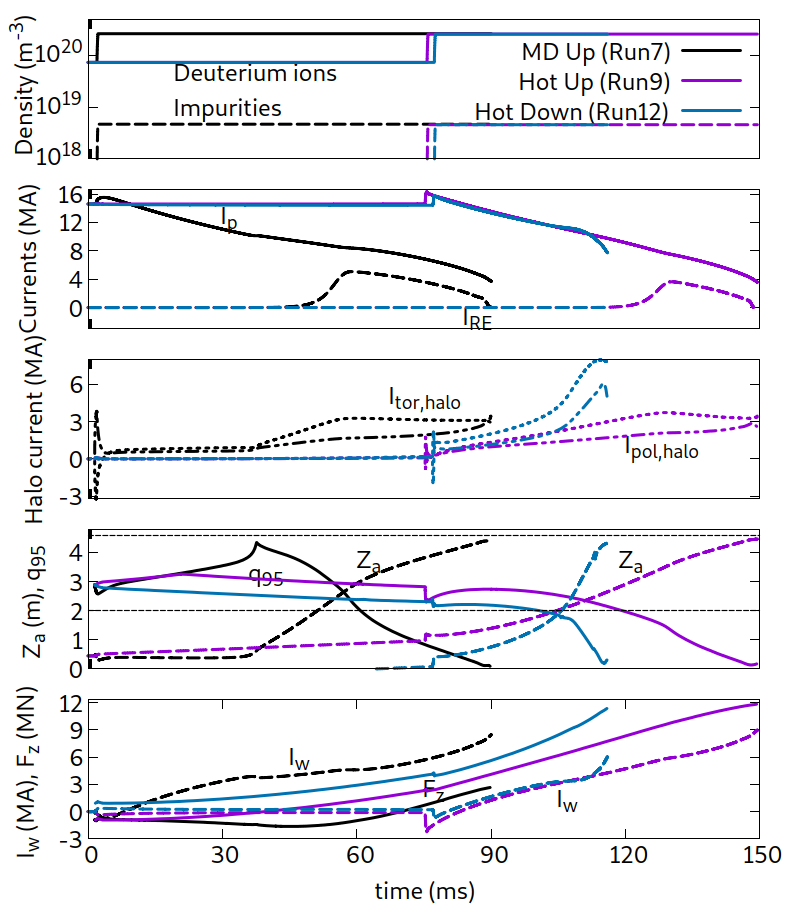}
\caption{Comparison of the evolution of 0D parameters comparing the baseline cases (without 2\textsuperscript{nd} injections) of major disruption (MD) (Run7) with that of the hot-VDE up (Run9) and hot-VDE down (Run12). Note that the sign for the wall force for the downward VDE case (Run12) has been reversed for easy comparison. }
\label{fig:md_up_down_all}
\end{figure}

The situation with the downward hot-VDE case (Run12) is however starkly different. Although the pre-TQ VDE growth rate is similar to that of the upward-hot-VDE (Run9), it gets much higher post the thermal quench. This can be seen from Fig.~\ref{fig:md_up_down_all} (4th panel) and Fig.~\ref{fig:md_up_down_Ip_Z}. This leads to a much faster shrinking of the plasma such that the effect of scraping-off dominates the RE avalanche, consequently leading to nearly no RE beam formation at all. In addition, in this situation, the plasma confined region vanishes at a significantly larger plasma current (see Fig.~\ref{fig:md_up_down_Ip_Z}). The faster timescale of vertical motion (and plasma shrinking) even at larger $I_p$ also leads to higher halo currents as can be seen from Fig.~\ref{fig:md_up_down_all}. Faster post-TQ vertical motion in the downward VDE can be attributed to the relatively stronger gradient in the magnetic field below the lower X-point in ITER, that in turn causes a stronger vertical force on the plasma during the downward VDE. Nevertheless, such a scrape-off dominated scenario could point to a potential route for the avoidance of RE beam formation albeit with the consequence of higher total current at final collapse and larger wall forces. Prior work \cite[]{DINA-report:2016} using DINA simulations of ITER downward hot-VDEs showed a similar behaviour, wherein the vertical motion and scrape-off was relatively far stronger than in upward VDEs. 

\subsection{Effect of 1\textsuperscript{st} injection Neon quantity}
\begin{figure}
\centering
\includegraphics[trim = 0mm 0mm 0mm 0mm, clip, width=0.65\textwidth]{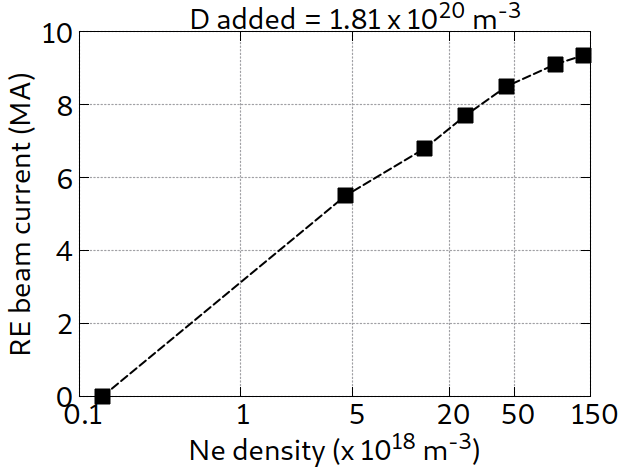}
\caption{Maximum RE beam current formed versus the 1\textsuperscript{st} injection quantity of Neon. The injected Deuterium was fixed at $n_\mathrm{D}=1.81\times 10^{20}\mathrm{m^{-3}}$.}
\label{fig:1st_Ne_sens}
\end{figure}

The effect of 1\textsuperscript{st} injection quantity of Ne on the RE beam current formed in MD cases based on Run1 is shown in Fig.~\ref{fig:1st_Ne_sens}. Injection of lower quantities of Neon (1\textsuperscript{st} injection) reduces the RE beam current formed. In fact, with a Ne injection quantity that is $3\%$ of that in Run1, it has been observed that there is no noticeable RE formation even after $100$ms (Run1a). This is due to the relatively quick and significant Ohmic reheating of the plasma in the absence of much Neon, post which there is not enough electric field available for an efficient RE avalanche. The present observation is in line with previous predictions with partially-ionized impurities \cite[]{Vallhagen:2020} and also with the usual absence of significant post-disruption RE beams in existing experiments without high-Z injections. However, with present assumptions in ITER, the requirement to radiate at least $\sim 85\%$ of thermal energy and to maintain a CQ time  $t_\mathrm{CQ}<150\mathrm{ms}$ would need a Neon 1\textsuperscript{st} injection density rise $\sim 10^{19}$ or higher (it is possible that the required rise can be slightly lower in case assimilation of H in 1\textsuperscript{st} injection can be higher).  This emphasises that the risk from REs in ITER is strongly enhanced by high-Z injections meant to simultaneously mitigate high thermal-loads on the divertor and high electromagnetic loads on the vacuum vessel. It must be noted however, that increasing the injected Ne beyond a certain level does not translate to a commensurate rise in the RE beam current (the extent of electric field available becomes important and limits the obtained RE beam current).

\subsubsection{Effect of Ne 1\textsuperscript{st} injection quantity on the overall evolution of MD and VDE cases}

As mentioned earlier, full simulations for MD (up), upward VDE and downward VDE have been performed also with a Neon 1\textsuperscript{st} injection quantity that is $30$ times larger. These simulations are referred to with the suffix 'L' in the table.~1. Here we compare the effect of the much larger Neon injection on the evolution of the disruption.

\begin{figure}
\centering
\includegraphics[trim = 0mm 0mm 0mm 0mm, clip, width=1.0\textwidth]{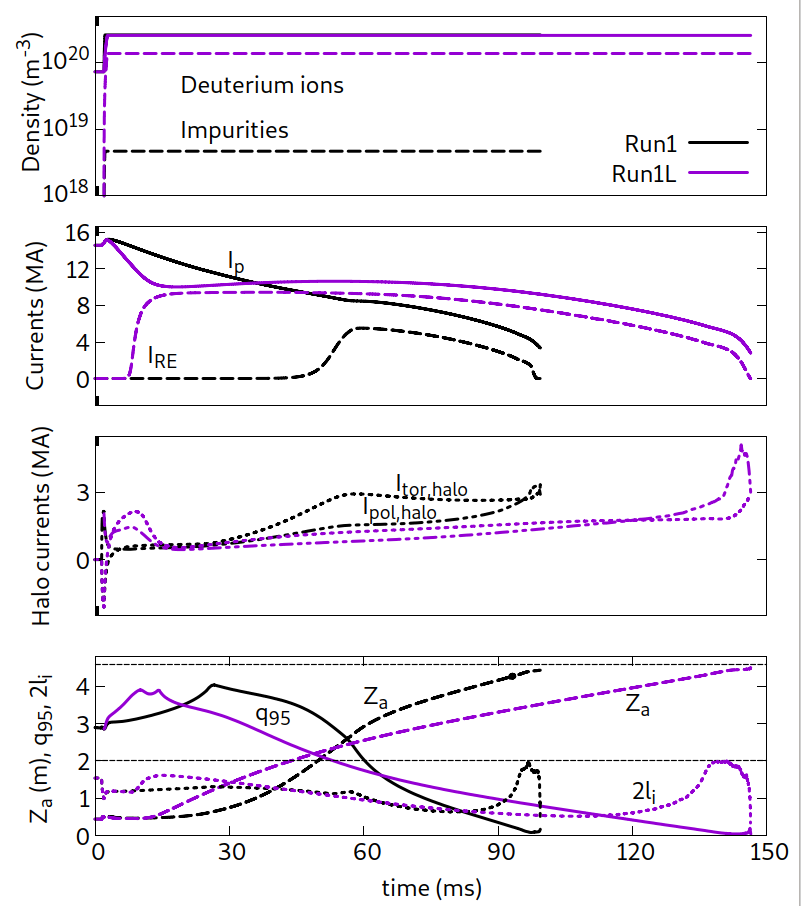}
\caption{Comparison of the evolution of 0D parameters for the baseline MD cases with different 1\textsuperscript{st} injection Ne quantities (Run1 and Run1L).}
\label{fig:run1_vs_run1L_all}
\end{figure}

\begin{figure}
\centering
\includegraphics[trim = 0mm 0mm 0mm 0mm, clip, width=0.65\textwidth]{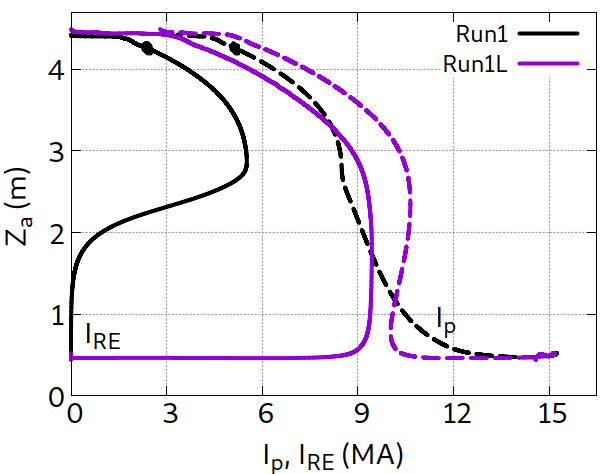}
\caption{Total plasma current and RE current versus the plasma axis vertical position (Run1 and Run1L).}
\label{fig:run1_vs_run1L_Ip_Z}
\end{figure}

The overall evolution is in many ways qualitatively similar even for the larger Ne injection. However, there are differences in details that arise. Figure.~\ref{fig:run1_vs_run1L_all} highlights the important differences in evolution of the global parameters for the MD baseline cases Run1 and Run1L. As expected, higher Ne quantity leads to a lower temperature after 1\textsuperscript{st} injection, and therefore a faster current quench. This causes the $q_{95}$ to peak much earlier since this is a phase when current decays without the plasma shrinking in size significantly ($q_{95} \sim a^2/I_p$). Faster CQ also causes the threshold $I_p$ value ($\sim 10$MA) to be reached earlier and hence the plasma vertical instability starts earlier in the case of larger Ne injection. Likewise, increase in the number of bound-electrons available for avalanche leads to a larger RE beam current $\sim 9.5$MA at higher Ne injection (Run1L) versus $\sim 5.5$MA for Run1. Significant plasma vertical motion and the associated halo current $\sim 3\mathrm{MA}$ already by the time of RE avalanching in Run1 also contributes to this difference in final RE current. However, more important is the decay rate of the RE beam current. In the case of Run1L, when the full RE beam current is formed, it is not immediately subjected to a decay via scrape-off unlike in Run1. Due to this the plateau phase is relatively flat and so the plasma motion toward the wall is much slower. Therefore, unlike in Run1, in the case of larger Ne injection, the decay of $q_{95}$ after peaking is a result of plasma shrinking at a nearly constant $I_p$. This can be seen clearly in Fig.~\ref{fig:run1_vs_run1L_Ip_Z}, in the vertical leg of the $I_p$ curves. However, what is practically important eventually is the fact that with larger Ne injection, due to relatively prompt conversion via RE avalanche, $q_{95}<2$ occurs far into the RE plateau phase. Differences in the overall evolution described here is quite similar in the case of upward VDE cases Run9 vs Run9L (not shown for the sake of brevity). 

\begin{figure}
\centering
\includegraphics[trim = 0mm 0mm 0mm 0mm, clip, width=1.0\textwidth]{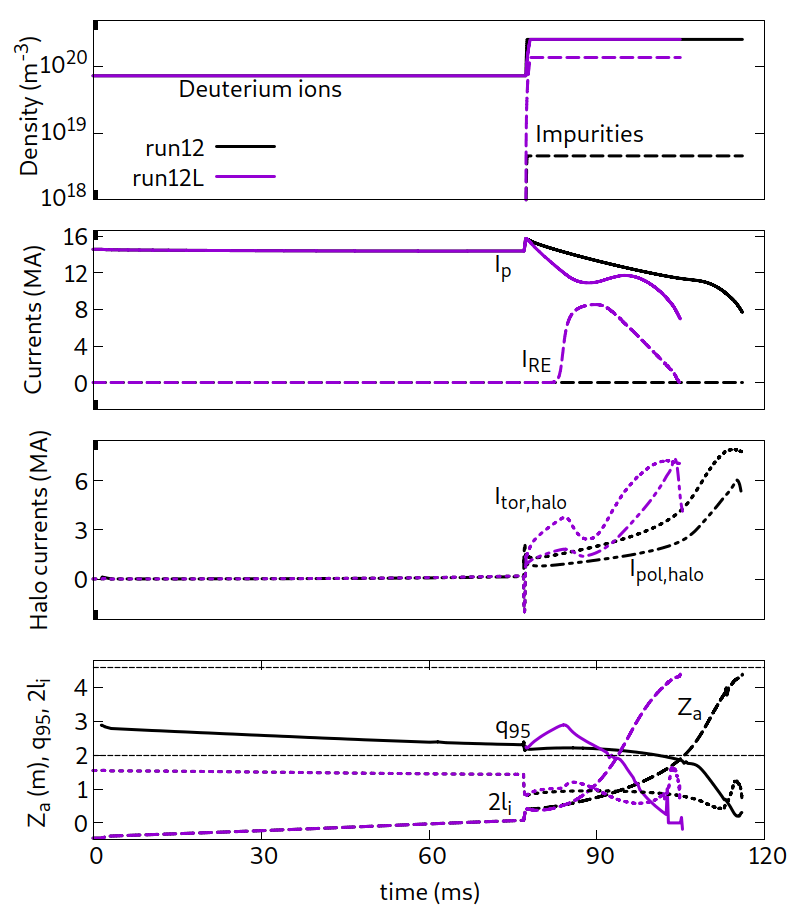}
\caption{Comparison of the evolution of 0D parameters for the baseline downward-VDE cases with different 1\textsuperscript{st} injection Ne quantities (Run12 and Run12L). Note that the label $Z_a$ in this specific case (downward motion) represents $-Z_a$.}
\label{fig:run12_vs_run12L_all}
\end{figure}

\begin{figure}
\centering
\includegraphics[trim = 0mm 0mm 0mm 0mm, clip, width=0.65\textwidth]{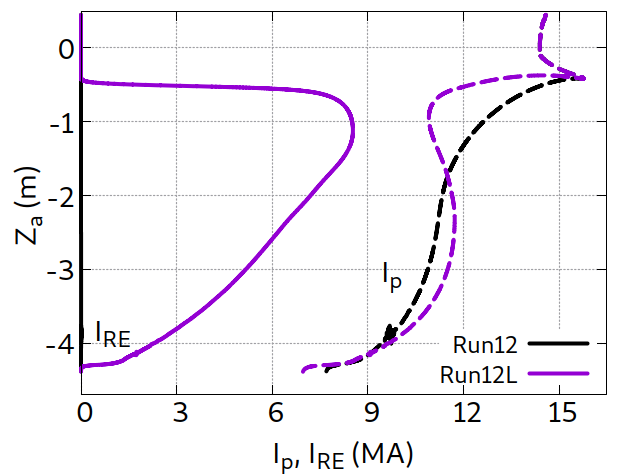}
\caption{Total plasma current and RE current versus the plasma axis vertical position (Run12 and Run12L).}
\label{fig:run12_vs_run12L_Ip_Z}
\end{figure}

However, in the case of downward VDE situation, the effect of larger Ne injection is different from upward VDs in some respects. This is shown in Fig.~\ref{fig:run12_vs_run12L_all}. At first, in spite of the generally much stronger vertical instability in downward VDs, with larger Ne injection, there is a significant RE beam formation ($\sim 8.5$MA in Run12L) as compared to a negligible beam formed with lower Ne injection (Run12). This is due to the fact that with higher Ne injection, the avalanche timescale is not too slow anymore as compared to the timescale over which the scrape-off causes it to decay. In the case of lower Ne injection, as mentioned earlier, before full avalanche growth occurs, scrape-off effect dominates and causes RE current decay. Nevertheless, unlike in the upward VDs (Run1L and Run9L), in the case of downward VDE (Run12L), the RE beam decays via scrape-off soon after reaching the plateau phase. One important consequence of this is the overall faster current decay in Run12L versus Run12, which causes the vertical motion timescale to be faster. In summary, larger Ne injection hastens the vertical plasma motion in a downward VDE, while it slows it down (on-average over the whole motion) in the case of upward VDs.

In terms of the effect of 2\textsuperscript{nd} Neon injection and Neon flushout, the overall evolution has been observed to be qualitatively similar for the large Ne-1\textsuperscript{st}-injection scenario as it was for the lower Ne-1\textsuperscript{st}-injection cases described earlier. Such an observation holds  for the MD cases (Run1L vs Run2L vs Run5L vs Run4L vs Run 6L), the upward VDE cases (Run9L vs Run10L vs Run11L) and downward VDE cases (Run12L vs Run13L vs Run14L) as well with large Ne 1\textsuperscript{st} injection (figures not shown for brevity). Larger RE beam currents with larger Neon 1\textsuperscript{st} injection also implies a relatively higher fraction of energy extracted by the REs. Also in terms of energy gain and dissipation by REs, the effect of 2\textsuperscript{nd} injection and Ne flushout was observed to be qualitatively similar to the cases with lower Ne 1\textsuperscript{st} injection. 

\section{Summary and outlook}

First axisymmetric simulations of ITER disruptions were presented that include runaway electrons, vertical plasma motion (VDEs) as well as massive material injections self-consistently. The work provides important insights into the interplay between RE formation, VDE and RE scrape-off, and timed massive material injections/flushout. Firstly, RE formation not only leads to additional cooling of the plasma but also slows down the decay of current during the disruption and hence delays the time to final collapse but at roughly similar final currents. While different levels of current-flattening can lead to differences in the disruption evolution, the effect in general is found to be significantly weaker due to the fast timescale current-profile-smoothing-out effect of the highly-resistive plasma. Due to this, for example, no noticeable difference was found between flattening within $q=2$ and a significant flattening within the LCFS. Given that high-fidelity 3D thermal-quench predictions for ITER can be challenging to obtain, the reliability of post-TQ predictions that that lack the input of a precise current-profile has often been considered questionable. However, the weak effect of the J-profile observed in our study indicates that in reality, a precise post-TQ current-profile might not be as important.

Neon 2\textsuperscript{nd} injection causes faster RE beam decay, but also a  correspondingly faster vertical motion to the wall. Effectively the RE current at final termination does not change significantly due to the 2\textsuperscript{nd} injection. While earlier injection (during avalanche) enhances the avalanche rate, the subsequent evolution remains similar to the plateau injection. On the other hand, Neon flushout not only has a far weaker effect on the overall dynamics but its timing doesn't make much difference either. Furthermore, even with Ne 2\textsuperscript{nd} injection or flushout, the undissipated RE energy is estimated to be $\sim70$ MJ (with lower Ne 1\textsuperscript{st} injection) and $\sim100-130$ MJ (with larger Ne 1\textsuperscript{st} injection), which would be eventually deposited via scrape-off and final collapse. The insensitivity of Ne 2\textsuperscript{nd} injection or flushout in this respect is due to the commensurate increase in the energy channeled to REs along with the energy dissipated. This means that 2\textsuperscript{nd} injection of Neon or flushout are not effective in reducing the total RE kinetic energy that gets dumped into the wall (of course 3D instabilities can change this picture). 

In general, the behaviour for the upward VDs (both MD and hot-VDEs) have been observed to be qualitatively very similar, except for differences caused by the time delay of TQ in the hot-VDE (and the consequent size and current density difference). One important difference in our simulations has been the attainment of $q_{95}<2$ in hot-VDEs at lower Neon 1\textsuperscript{st} injection earlier than any significant RE beam forms. This opens up the \emph{possibility} for a benign situation of instability-induced stochastic plasma causing partial/full loss of remnant REs (primary seed and partly exponentiated REs). However, with larger Neon 1\textsuperscript{st} injection, attainment of $q_{95}<2$ occurs only after the full RE beam is formed.

We observe that the downward VDE is strongly unstable post the thermal-quench such that the plasma reaches the final collapse significantly earlier and with significantly larger current. Faster gradients in the magnetic field below the x-point is the general reason for the faster timescale of downward VDEs. One of the consequences is that the scrape-off doesn't allow any significant avalanche growth in the case of low-Ne-1\textsuperscript{st}-injection (Run12) such that no RE beam is formed.  However, with larger Ne-1\textsuperscript{st}-injection, a multi-MA RE beam still forms before promptly decaying via scrape-off.

Larger Neon 1\textsuperscript{st} injection quantities clearly lead to larger final RE beam currents. The fundamental difference (w.r.t large Neon-1\textsuperscript{st}-injection) in the overall evolution of the disruption occurs due to the timescale changes in the vertical motion and RE avalanche. Due to this, for upward VDs, we observe a flat RE plateau phase (as compared to being scrape-off dominated) and an overall slowdown of vertical motion at larger Ne 1\textsuperscript{st} injection. For downward VDEs this causes a multi-MA RE beam and also a faster vertical motion. Neon 1\textsuperscript{st} injection quantity doesn't however change the qualitative evolution of the disruptions w.r.t the effect of Neon 2\textsuperscript{nd} injection and Neon flushout.

Finally, it must be pointed out that the relative timescale in ITER of Neon flushout as compared to the time for the RE beam plateau to final collapse is an important aspect that is often overlooked when extrapolating the benefits of present-day benign terminations to ITER. The typical time for RE beam decay observed in our simulations is $\sim 40$ms. So in case Neon recombination and outward transport of Neon neutrals (flushout) in ITER occurs at a comparable timescale, then the expected benign effects of flushout (as observed in present day experiments) might turn out to be negligible or at most marginal. This calls for an urgent and reliable prediction of the recombination and flushout timescales in ITER. Finally, in the scenario where Ne flushout doesn't lead to benign effects, faster scrape-off loss of REs or edge safety factor drop (and associated stochastic and distributed RE loss) might still hold a possibility for a benign termination. Investigation of such benign terminations in ITER with 3D MHD simulations using JOREK has been performed and is outside the scope of the present paper \cite[]{Bandaru-NF:2024}.

\section*{Acknowledgements}
The authors acknowledge fruitful discussions with E. Nardon. This work has been carried out within the framework of the ITER implementing agreement No.3, Ref:IO/IA/20/4300002200. ITER is the Nuclear Facility INB No.~174. This work explores the physics processes during plasma operation of the tokamak when disruptions take place; nevertheless the nuclear operator is not constrained by the results presented here. The views and opinions expressed herein do not necessarily reflect those of the ITER Organization. Part of this work was supported by the EUROfusion—Theory and
Advanced Simulation Coordination (E-TASC) via the theory and simulation verification and validation (TSVV) project on MHD transients (2021-2025). Part of this work has been carried out within the framework of
the EUROfusion Consortium, funded by the European Union via the Euratom Research
and Training Programme (Grant Agreement No. 101052200 of EUROfusion).  Views and opinions expressed are however those of the author(s) only and do not necessarily reflect those of the European Union or the European Commission. Neither the European Union nor the European Commission can be held responsible for them.

\newpage
\bibliographystyle{jfm}
\bibliography{2D_ITER_sims}

\end{document}